\documentclass[conference]{IEEEconf}

\input epsf
\usepackage{graphicx}
\usepackage{multirow}
\usepackage{titlesec}
\usepackage{balance} 
\usepackage{stfloats}
\usepackage{xcolor}
\usepackage{cite}
\usepackage[colorlinks,
            linkcolor=black,       
            anchorcolor=black,
            citecolor=black,      
            ]{hyperref}
\usepackage[normalem]{ulem}
\useunder{\uline}{\ul}{}
\usepackage[numbers,sort&compress]{natbib}
\usepackage{enumitem}
\usepackage{diagbox}

\renewcommand\thesection{\arabic{section}} 

\renewcommand\thesubsectiondis{\thesection.\arabic{subsection}}

\renewcommand\thesubsubsectiondis{\thesubsectiondis.\arabic{subsubsection}}

\begin{document}

\title{\textbf{\Large A study on the impact of pre-trained model on Just-In-Time defect prediction\\}}

\author{Yuxiang Guo$^{1}$, Xiaopeng Gao$^{2}$, Zhenyu Zhang$^{3}$, W.K.Chan$^{4}$, Bo Jiang$^{5,*}$\\
	\normalsize $^{1,2,5}$State Key Laboratory of Software Development Environment, School of Computer Science
and Engineering, \\
Beihang University, Beijing, China \\
	\normalsize $^{3}$State Key Laboratory of Computer Science Institute of Software, Chinese Academy of Sciences, Beijing, China \\
	\normalsize $^{4}$Department of Computer Science, City University of Hong Kong, Hong Kong\\
	\normalsize irisg@buaa.edu.cn, gxp@buaa.edu.cn, zhangzy@ios.ac.cn, wkchan@cityu.edu.hk, jiangbo@buaa.edu.cn \\
 \normalsize $^{*}$corresponding author \\
}


\maketitle
\begin{abstract}
Previous researchers conducting Just-In-Time (JIT) defect prediction tasks have primarily focused on the performance of individual pre-trained models, without exploring the relationship between different pre-trained models as backbones. In this study, we build six models: RoBERTaJIT, CodeBERTJIT, BARTJIT, PLBARTJIT, GPT2JIT, and CodeGPTJIT, each with a distinct pre-trained model as its backbone. We systematically explore the differences and connections between these models. Specifically, we investigate the performance of the models when using Commit code and Commit message as inputs, as well as the relationship between training efficiency and model distribution among these six models. Additionally, we conduct an ablation experiment to explore the sensitivity of each model to inputs. Furthermore, we investigate how the models perform in zero-shot and few-shot scenarios. Our findings indicate that each model based on different backbones shows improvements, and when the backbone's pre-training model is similar, the training resources that need to be consumed are closer. We also observe that Commit code plays a significant role in defect detection, and different pre-trained models demonstrate better defect detection ability with a balanced dataset under few-shot scenarios. These results provide new insights for optimizing JIT defect prediction tasks using pre-trained models and highlight the factors that require more attention when constructing such models. Additionally, CodeGPTJIT and GPT2JIT achieved better performance than DeepJIT and CC2Vec on the two datasets respectively under 2000 training samples. These findings emphasize the effectiveness of transformer-based pre-trained models in JIT defect prediction tasks, especially in scenarios with limited training data. 
\end{abstract}
\IEEEoverridecommandlockouts
\vspace{1.5ex}

%
\IEEEpeerreviewmaketitle

\section{Introduction}
The existence of software defects may lead to significant economic loss to the end users. Defect prediction can help developers identify defects before deploying buggy software.Just-In-Time(JIT for short) defect prediction predicts the presence or absence of defects based on software change characteristics\cite{2008Classifying}.

Feature extraction using deep learning technology has become a common method for deep JIT defect prediction models like DeepJIT\cite{8816772} and CC2Vec\cite{9284081}. DeepJIT is an end-to-end JIT defect prediction model that predicts defects with the Commit code and Commit message in a submission\cite{8816772}. CC2Vec improves DeepJIT by a pre-trained distributed change-code vector in advance, resulting in performance improvements on both the QT and Openstack\cite{9284081}.

Hong\cite{9284081} believes that the extra pre-trained code vector is an important reason for the improved performance of CC2Vec model in JIT defect prediction. For the traditional two-step deep learning models based on train-test such as DeepJIT, they only learn domain knowledge from scratch without any external knowledge. Different from DeepJIT, CC2Vec utilizes pre-trained code vector as a part of the input in JIT defect prediction model.

However, CC2Vec introduces extra information before training a model, that information is still intra-domain which means all the information learned by DeepJIT and CC2Vec are from the same dataset. The appearance of pre-trained models provides deep learning methods with a new way of pre-training and fine-tuning to resolve the previous tasks. Some transformer encoder-based pre-trained models like RoBERTa\cite{2018BERT}, CodeBERT\cite{2020CodeBERT}, decoder-based pre-trained models like GPT2\cite{Radford2019LanguageMA}, CodeGPT, encoder-decoder based model like BART\cite{lewis2019bart}, PLBART\cite{ahmad2021unified}, T5\cite{T5} have shown huge improvement in previous natural languages tasks. Benefit from the emergence of pre-trained models some researchers improve program language tasks like clone detection, code search and vulnerability detection to construct a pre-trained model based framework\cite{9609166,msr22}. Zhou\cite{9609166} evaluated the performance of CodeBERT in JIT defect prediction, showing that CodeBERT can achieve consistent performance with the current optimal model CC2Vec. But one problem is that most of the program lanauage related work is based on CodeBERT model but few works have explored how transformer-based pre-trained models perform in JIT defect prediction tasks other than CodeBERT. Which means they tend to use transformer encoder-based model without considering other types. Therefore, it is crucial to systematically explore how different models behave and what they are sensitive to.

In this study, we build six deep JIT prediction models: RoBERTaJIT, CodeBERTJIT, GPT2JIT, CodeGPTJIT, BARTJIT, and PLBARTJIT, using corresponding pre-trained models and CNN network. To comprehensively evaluate these pre-trained JIT defect prediction models, we selected three transformer-based models and their natural language and programming language versions as backbones. Our primary motivations were to investigate the performance of different transformer structures in JIT defect prediction tasks and how they behave in different situations. We used Commit code and Commit message as model inputs and evaluated the models based on metrics such as AUC score, accuracy, precision, recall, and F1 score to compare their performance. Additionally, we compared the training efficiency and distribution similarity of prediction results among the six JIT prediction models. We also conducted ablation experiments on different input parts of the four models to analyze the influence of input data on their performance. Lastly, we explored the models' performance under zero-shot and few-shot scenarios. Our implementation of experiments is accessible at \href{ https://github.com/AresXD/JIT_defect_prediciton}{\textcolor{black}{https://github.com/AresXD/JIT\_defect\_prediciton}}.

The contribution of this work is fourfold.
\begin{itemize}
    \item First, rather than concentrating solely on individual model performance, our research delves deeply into the comparative analysis of these models based on different backbones.
    \item Second, we analyze the correlation between training efficiency and model distribution across the six models and we find that when the backbone's pre-training model is similar, the training resources that need to be consumed are much closer.
    \item Third, we conduct an extensive ablation experiment to gain insights into the sensitivity of each model to various inputs and find that Commit code plays a significant role in defect detection between Commit code and Commit message, which provides the idea of optimization: researchers can focus on the processing of Commit code. 
    \item  Finally, we explore the performance of different models under zero-shot and few-shot learning scenarios. We find that the pre-trained models demonstrate superior defect detection abilities with a balanced dataset, and we obtain two models which can outperform DeepJIT and CC2Vec under a 2000-shot scenario, which provides the possibility to optimize a model under the few-shot scenario.
\end{itemize}
\section{Background on JIT defect prediction and pre-trained model }
In this section, we will revisit background information on JIT defect prediction and pre-trained models.
\subsection{Definition of submission in JIT defect prediction}
\begin{figure}[h]
    \centering
    \includegraphics[width=0.48\textwidth ]{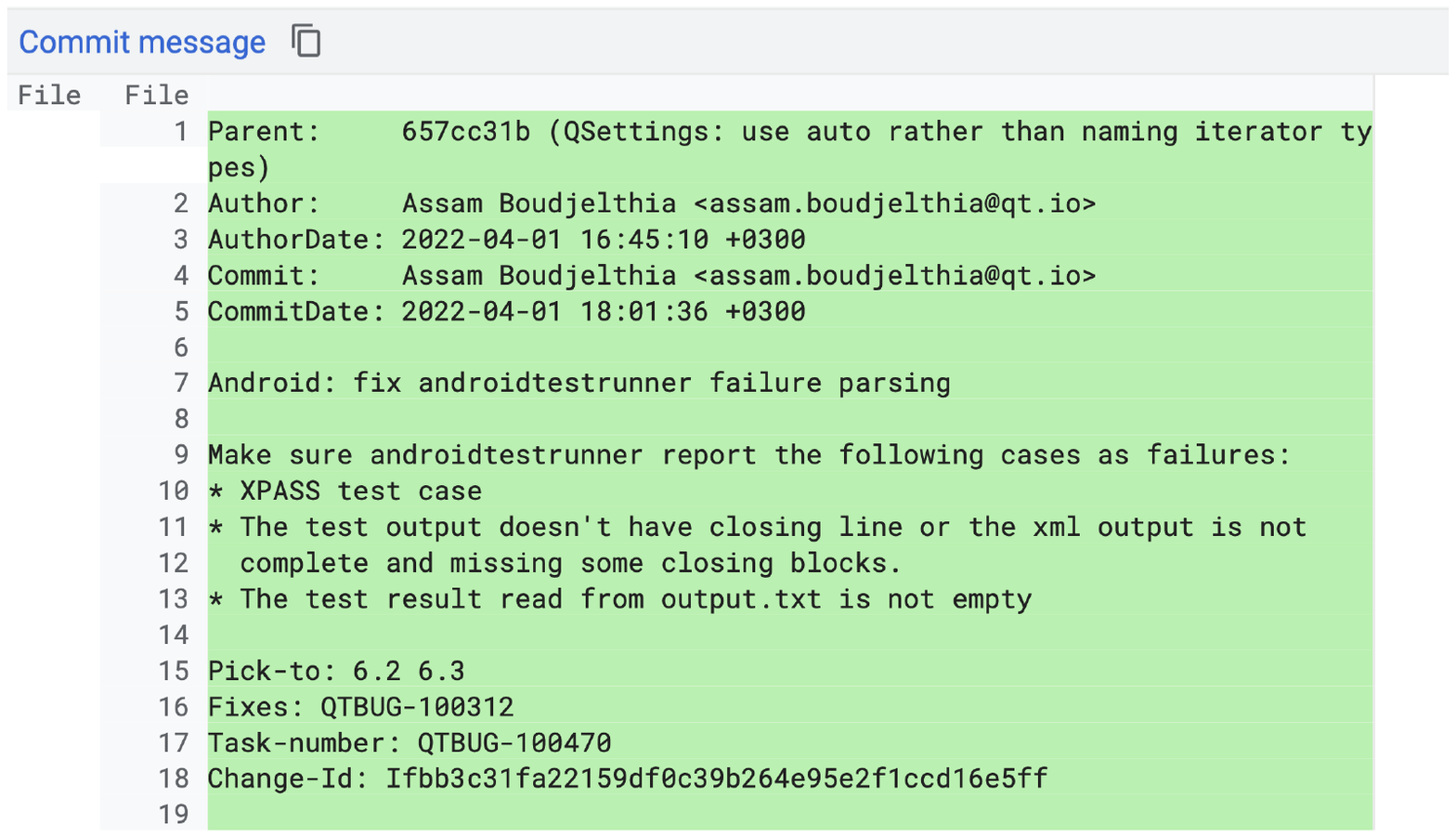}
    \caption{An example Commit message}
    \label{fig:f2}
\end{figure}
\begin{figure}[h]
    \centering
    \includegraphics[width=0.48\textwidth ]{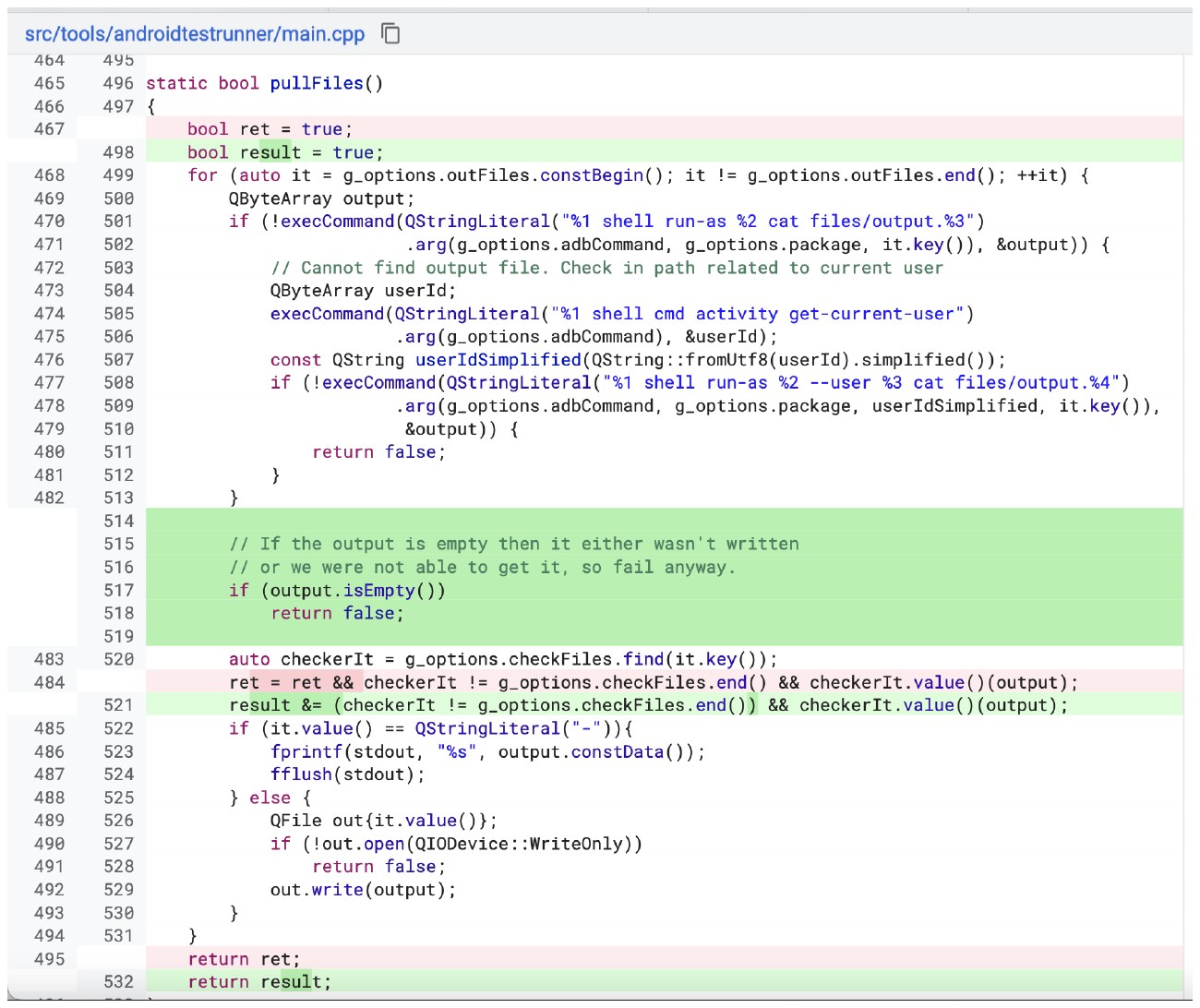}
    \caption{An example Commit code}
    \label{fig:f3}
\end{figure}
Developers often modify more than one file when submitting code to a project. Figure \ref{fig:f2} presents an example of a Commit message while Figure \ref{fig:f3} presents the example of a Commit code within one submission. The Commit message shows the developer who submitted the code (line 4), the commit time (line 5), the task number (line 17), and the change ID (line 18). The most important information in the Commit message is the developer's annotation ( lines 7-13). The developer's annotation is made up of natural language text that reflects the developer's intention for submission. Annotation is a useful feature often used in deep JIT defect prediction tasks, which is one of the input data used in our model. For convenience, we call the annotated data ``Commit message''.

Figure \ref{fig:f3} shows the part of the changed code in a submission whose changed ID is ``Ifbb3c31fa22159df0c39b264e95e2f1ccd16e5ff''. The green and red code identifies changed code in the file whose path within the project is ``src/tools/androidtestrunner/main'', the green part represents the added code and the red part represents the deleted code. In Figure \ref{fig:f3}, we can see that the major bug of this submission is replacing the variable ``ret'' with ``result'' and changing the symbol ``\&\&'' to ``\&''. From Figure \ref{fig:f3}, we know that there are two types of changed code in a submission, added code and deleted code. Changed code in a submission is important information, which is another part of the input data used in our model. For convenience, we also call the changed code ``Commit code''.
\subsection{DeepJIT and CC2Vec}
Hong et al. introduced a novel deep Just-In-Time (JIT) defect prediction model, utilizing Commit code and Commit message as the model input. In contrast to previous approaches that relied on traditional handcrafted features, DeepJIT represents an end-to-end JIT defect prediction model, eliminating the need for manual feature extraction\cite{8816772}. The model utilizes a Convolutional Neural Network (CNN) to extract meaningful features to achieve classification tasks.

To find a suitable vector representation with generalization, Hong et al.\cite{9284081} propose a neural network model that can learn a representation of code changes by pre-training the Commit code into a distributed vector CC2Vec in advance. CC2Vec uses HAN\cite{2017Multilingual} network architecture to better extract semantic information from code by training vectors using information from Commit messages submitted by developers at a time. Hong et al. achieved performance improvements in the JIT defect prediction task by concatenating the CC2Vec vector with other input data from DeepJIT.
\subsection{Pre-trained model}
Word embedding or distributed vector is a technique for converting natural language words into vectors or matrices such that computers can process and manipulate. Only by using reasonable word representation can we carry out subsequent machine learning more easily. The development of a pre-trained model is closely related to the research of word embedding. The first generation of pre-trained models is pre-trained word embedding. Some models, such as one-hot, Word2Vec\cite{2013Distributed}, N-Gram\cite{2013Efficient} and Glove\cite{2014Glove}, can capture certain semantic information. But they are context-free. The first generation of pre-trained models has obvious shortcomings. For example, the embedding is static and is unable to consider the context, semantic role, and syntactic characteristics of the text, which makes it hard to distinguish polysemous words effectively. Therefore, another stage of the development of word embedding is to study context-related word embedding.

\begin{figure}[h]
    \centering
    \includegraphics[width=0.44\textwidth ]{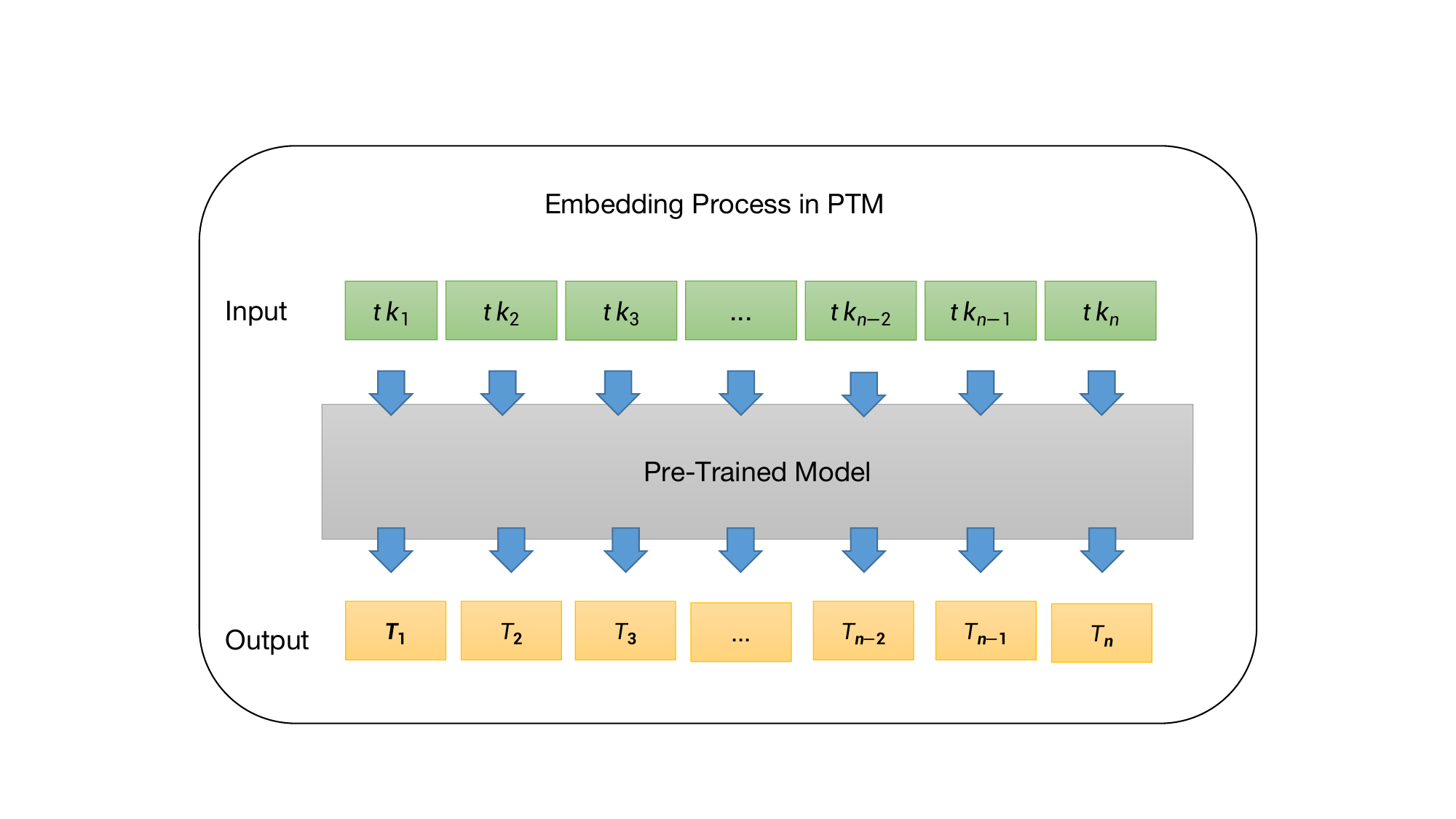}
    \caption{The process of using Pre-trained models}
    \label{fig:f4}
\end{figure}

The second generation of the pre-trained model depends on the proposed transformer structure\cite{2017Attention}. The main pre-trained models fall into three categories: encoder based such as ELMo\cite{2018DeepC}, BERT\cite{2018BERT}, and RoBERTa\cite{2019RoBERTa}, decoder based such as GPT\cite{radford2018improving}, GPT2\cite{Radford2019LanguageMA}, encoder-decoder based such as BART\cite{lewis2019bart}. These pre-trained models are usually trained on massive corpora and support many training tasks. Large-scale pre-training can learn more general language representation, provide a better initialization model for downstream tasks, improve the performance of target tasks, and accelerate the convergence of models. By continuously improving the existing model structure and adding code-related corpora, researchers have derived code-related versions of the corresponding pre-trained models like CodeBERT, CodeGPT, PLBART, etc. Which models obtain great performance in clone detection, vulnerability detection\cite{9609166,msr22}. But for JIT defect prediction task, the role of these models, and the scenarios in which they work, remains unexplored. Figure \ref{fig:f4} shows the process of using a pre-trained model that feeds a set of tokens into the pre-trained model and outputs a generalized language representation vector. Finding suitable word embedding has been the focus of natural language research for a long time. Code is a semi-structured language similar to natural language, so when we use deep learning to solve problems related to program language, it is also necessary to find an appropriate embedding accordingly.
\section{Model Structure}
In our research, we developed six distinct JIT defect prediction models by leveraging pre-trained models as backbones and employing CNN networks for feature extraction. Specifically, we incorporated three transformer-based models: RoBERTa (an encoder-based model), GPT2 (a decoder-based model), and BART (an encoder-decoder-based model) to create RoBERTaJIT, GPT2JIT, and BARTJIT, respectively. Additionally, we extended the scope by selecting three code-related versions from the RoBERTa, GPT2, and BART models to establish CodeBERTJIT, CodeGPTJIT, and PLBARTJIT models. These models were designed to explore the effectiveness of different transformer structures in the JIT defect prediction task, aiming to provide comprehensive insights into their performances and capabilities.

\begin{figure}[h]
    \centering
    \includegraphics[width=0.44\textwidth ]{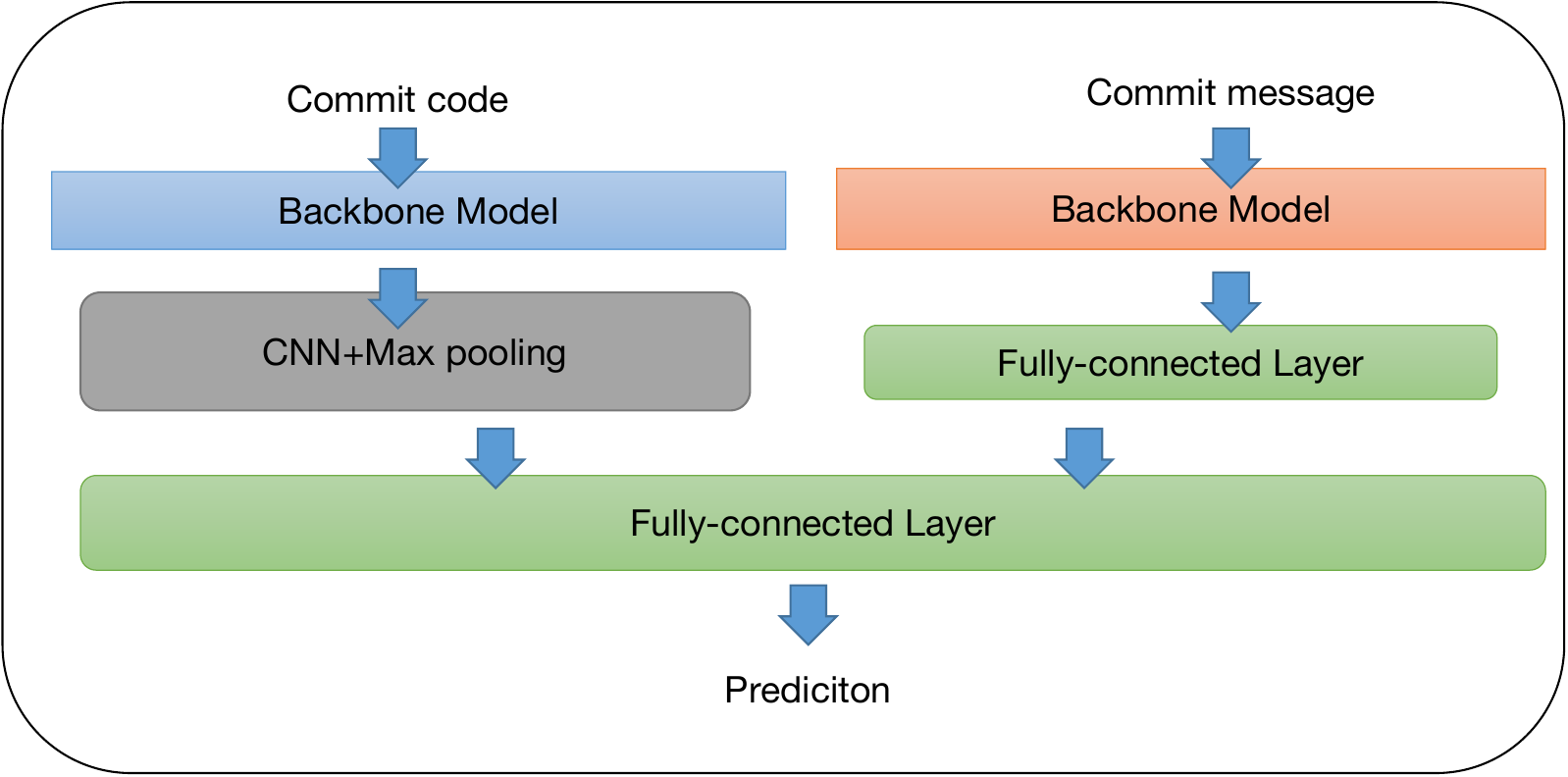}
    \caption{The framework of JIT defect prediction models with different pre-trained model as backbones}
    \label{fig:model}
\end{figure}
Figure \ref{fig:model} illustrates the framework of our six JIT defect prediction models, all of which utilize Commit code and Commit message as the model input. In the process of handling the Commit code, we adopt a similar approach as the DeepJIT model\cite{8816772}, where we select the most recent four patches from each submission. The Commit code, denoted as $C=[P_1,P_2,...,P_{|C|}]$, comprises $|C|$ patches in a submission, where $|C|=4$ represents the number of patches. In cases where the number of patches in a submission is less than 4, we employ "empty patches" to pad the remaining slots on the left. Each individual patch $P_i$ is then treated as a sequence within the Commit code, which serves as input to our backbone pre-trained model for obtaining the embedding representation $\bar{P_i}$ using Equation (1).
\begin{eqnarray}
\bar{P_i} =BackboneModel(P_{i}) 
\end{eqnarray}

When Commit code in each submission is processed by backbone pre-trained model, we obtain the vector representation of Commit code $\bar{C}$
using $\oplus$ operation following Equation (2).
\begin{eqnarray}
\bar{C}=\bar{P_1}\oplus{\bar{P_2}}\oplus{...}\oplus{\bar{P_{|C|}}}
\end{eqnarray}

For the process of Commit message, we treat the whole Commit message as sequence $M=[w_1,w_2,w_3,...w_{|m|}]$ in a submission whose sequences content contains 13-17 lines in Figure \ref{fig:f2}, $M$ is a sequence composed of a set of words, and $|m|$ is the length of the sequence. Then we input Commit message $M$ into the backbone model to get embedding representation $\bar{M}$ following Equation (3).
\begin{eqnarray}
\bar{M} =BackboneModel(M) 
\end{eqnarray}

Then we conduct feature extraction for $\bar{C}$ and $\bar{M}$ respectively. We use CNN and max-pooling to extract the features of code embedding $\bar{C}$. We define a filter $f \in R^{k\times d}$ and use $K$ windows to extract new features of the Commit code. For the code patch $\bar{P_i}$ in $\bar{C}$, we obtain new vector $x_i$ following Equation (4). 
\begin{eqnarray}
x_i=ReLU (f*\bar{P_{i:i+k-1}}+b_i)
\end{eqnarray}

In this paper, symbol $*$ is the sum of the product of the element, $ReLU$ is the nonlinear activation function and $b_i \in R$ is the bias value. New features are generated by applying filter $f$ to every $k$ file of the Commit code. The new feature $x_i$ is strung together into vector $X$, as shown in Equation (5). 
\begin{eqnarray}
X=[x_1,x_2,...,x_{|C|-k+1}]
\end{eqnarray}

Then we use the max-pooling method to process vector $X$ to obtain the most significant feature $Z_c$, as shown in Equation (6).
\begin{eqnarray}
Z_c = \mathop{max}\limits_{1 \leq i \leq |C|-k+1}\ (x_i)
\end{eqnarray}

Because the sequence length of the Commit message is shorter than that of the Commit code, only one layer of a fully connected network is used for feature extraction in message embedding $\bar{M}$. We can obtain feature vector $Z_m$ following Equation (7).
\begin{eqnarray}
Z_m=w_m\cdot \bar{M}+b_m
\end{eqnarray}

After obtaining the feature vector $Z_c$ for the Commit code and $Z_m$ for the Commit message, we fuse the $Z_c$ and $Z_m$ to get the feature vector $Z$ for one submission, as shown in Equation (8).
\begin{eqnarray}
Z=Z_c \oplus Z_m
\end{eqnarray}
Then we input $Z$ into a layer of a fully connected network as a classifier. Finally, the high-dimensional vector is mapped to the one-dimensional vector by the $sigmod$ function to complete the prediction task of binary classification, as shown in Equation (9).
\begin{eqnarray}
y=sigmod(ReLU(w_h \cdot Z+b_h)) 
\end{eqnarray}
\section{Experimental research}
\subsection{Research Questions}
{\bfseries RQ1}: What is the performance of six pre-trained based JIT defect prediction models with different backbone structures?

{\bfseries RQ2}: What is the relationship between the training effectiveness of different models and the distribution of predicted results?

{\bfseries RQ3}: How sensitive are the different models to the input data?

{\bfseries RQ4}: How do different models behave in few-shot scenarios?

\subsection{Datasets}
All of our experiments are carried out on QT and Openstack datasets. This dataset was originally collected and cleaned by McIntosh and Kamei and come from two famous open source project QT and Openstack\cite{dataset2018}.  Table \ref{tab:dataset} shows the statistical description of QT and Openstack dataset. There are 25704 submissions in the QT dataset, including 1825 submissions that are labeled as defective. There are 1627 submissions in the Openstack dataset, including 1627 submissions that are labeled as defective. In this paper, 80\% of the data is used as the training set, and the left 20\% of the data is used as the test set for evaluation.
\begin{table}[h]
\renewcommand{\arraystretch}{1.5}
\tabcolsep=0.4cm
\caption{Statistics of the Dataset}
\label{tab:dataset}
\begin{center}
\small
\begin{tabular}{|c|c|c|c|}
\hline
{\bf{Project}} & \bf{Commit} & \bf{Defect} & \bf{\%Defect} \\ \hline
QT           & 25704              & 1825               & 7.1                \\ \hline
Openstack    & 13304               & 1627               & 12.2                 \\ \hline
\end{tabular}
\end{center}
\end{table}\vspace{-1.5em}
\subsection{Experiment Setup}
We use a desktop PC as our experiment environment. The PC is running Ubuntu 18.04 and is equipped with RTX3090 with 24GB of memory. All experiments are implemented by Python3 and we use the official Python release version 3.7. We select the set of parameters that can minimize the loss of the validation set, the corresponding model is saved as the final model in our training process.
\subsection{Criteria for evaluation}
When a binary model is used for prediction, four different results are produced. A commit that introduces a defect is predicted to be defective namely True Positive(TP), a commit that does not introduce a defect was predicted to be defect-free namely True Negative(TN), a commit that does not introduce a defect was predicted to be defective namely False Positive(FP) and a commit which introduces defect was predicted to be defect-free namely False Negative(FN).
\begin{figure*}
    \centering
    \includegraphics[width=0.90\textwidth ]{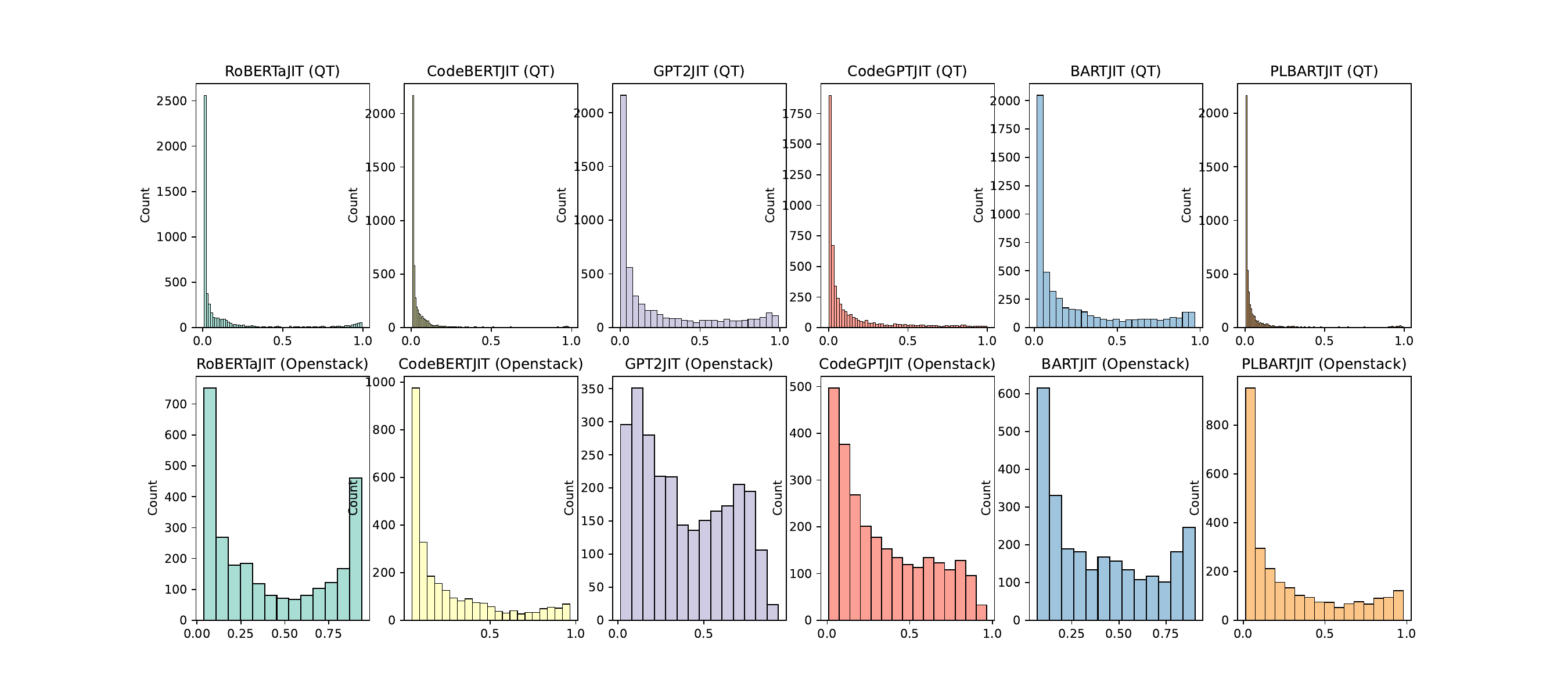}
    \caption{Distribution of prediction results for different models in QT and Openstack datasets}
    \label{fig:QT_op_distribution}
\end{figure*}
{\bfseries AUC} Area under the Curve of ROC(AUC) is often used to evaluate the training effect of a binary model. A high AUC score shows off the stability of a model.

{\bfseries Accuracy} In particular, Accuracy (ACC) is the ratio of the number of correct predictions over the total number of predictions. 
$$ACC=\frac{TN+TP}{TN+TP+FN+FP}$$

{\bfseries Precision} Precision rate refers to the proportion of correctly predicted defects among all predicted defects.
$$Precision=\frac{TP}{TP+FP}$$

{\bfseries Recall} Recall refers to the proportion of defects correctly predicted to all defects submitted. 
$$Recall=\frac{TP}{TP+FN}$$

{\bfseries F1 score} F1 score is a metric used to evaluate the performance of a binary classification model. It is the harmonic mean of precision and recall, and it takes both false positives and false negatives into account. The F1 score is a value between 0 and 1, where 1 represents perfect precision and recall, and 0 represents the worst possible performance.
$$F1=\frac{2*Precision*Recall}{Precision+Recall}$$
\subsection{Results and Discussions}
{\bfseries Results and analysis for RQ1.} For this research question, we try to explore the overall performance of different models in JIT defect prediction task under full fine-tuning scenario. To achieve this, we compare six models, each utilizing different backbone pre-trained models: RoBERTaJIT, CodeBERTJIT, GPT2JIT, CodeGPTJIT, BARTJIT, and PLBARTJIT. Additionally, we include two state-of-the-art models, DeepJIT and CC2Vec, for comparison. Most of the previous work only select auc score as evaluation index for JIT defect prediction model, we acknowledge that relying on a single metric may be limiting. To comprehensively explore the performance of these models, we additionally considered other evaluation metrics in this section and performed results analysis. Table \ref{tab:full-finetuneres-QT} and Table \ref{tab:full-finetuneres-op} shows the experimental results of these six models on QT and Openstack projects. 
\begin{table}[h]
\renewcommand{\arraystretch}{1.5} 
\tabcolsep=0.06cm
\caption {Experimental results of the different models on 
QT project}
\label{tab:full-finetuneres-QT}
\scalebox{.6}

\par\centerline{
\small
\begin{tabular}{|c|c|c|c|c|c|}
\hline
{\bf{Model}}&  auc score &  accuracy &  precision &  recall &   F1 score \\ \hline
DeepJIT &       0.79 &      0.73 &       0.17 &    0.70 & 0.27 \\ \hline
CC2Vec &       0.81 &      0.56 &       0.13 &    0.87 & 0.23 \\ \hline
 RoBERTaJIT &       0.81 &      0.87 &       0.27 &    0.48 & 0.35 \\ \hline
CodeBERTJIT &       0.81 &      0.87 &       0.26 &    0.42 & 0.32 \\ \hline
    GPT2JIT &       0.81 &      0.81 &       0.21 &    0.58 & 0.31 \\ \hline
 CodeGPTJIT &       0.81 &      0.89 &       0.29 &    0.39 & 0.33 \\ \hline
    BARTJIT &       0.82 &      0.82 &       0.23 &    0.62 & 0.34 \\ \hline
  PLBARTJIT &       0.82 &      0.90 &       0.32 &    0.34 & 0.33 \\ \hline
\end{tabular}}
\end{table}

As shown in Table \ref{tab:full-finetuneres-QT} and Table \ref{tab:full-finetuneres-op}, all models with various pre-trained models yield significant enhancements across the evaluation metrics, with the exception of recall. It is clear that decoder-based models and encoder-decoder-based models are typically well-suited for generation tasks. However, in the JIT defect prediction task, which involves understanding, we observe that pre-trained models deliver excellent performance regardless of their different architectures. This suggests that when the model size is sufficiently large, it demonstrates adaptability to tasks that may not be inherently suited to its architecture.
\begin{itemize}
    \item Results 1: All JIT defect prediction models constructed with three different transformer structures exhibit improvements in both datasets.
\end{itemize}

\begin{figure*}[h]
    \centerline{\includegraphics[width=0.8\textwidth ]{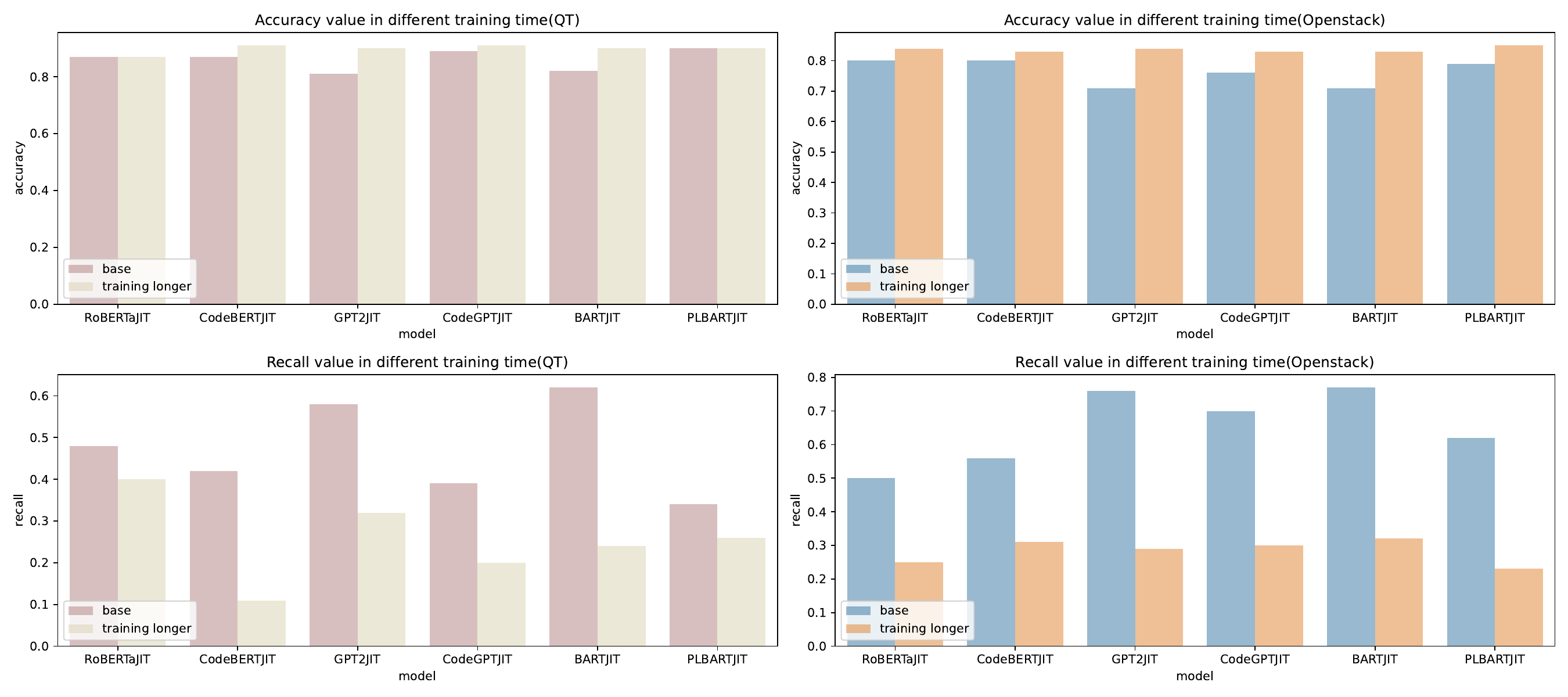}}
    \caption{Variation of accuracy and recall with training time on QT and Openstack datasets}
    \label{fig:base_longer}
\end{figure*}

Figure \ref{fig:QT_op_distribution} illustrates the distribution of prediction results for the six models on both the QT and Openstack datasets. It is evident that the models tend to predict most submissions as non-defective commits in the QT dataset, whereas this behavior differs in the Openstack dataset. This observation is also reflected in the results presented in Table \ref{tab:full-finetuneres-QT} and Table \ref{tab:full-finetuneres-op}, where the recall value for the six models on the Openstack dataset significantly outperforms that on the QT dataset. While the difference in recall performance in the previous two state-of-art models, DeepJIt and CC2vec, is very small. To further investigate this phenomenon, we compared the existing results with the results obtained after extending the model's training time. Figure \ref{fig:base_longer} depicts the differences between the base results and those achieved with longer training times. It is apparent that with increased training time, the accuracy of all six models improves, while the recall value deteriorates. This observation can be attributed to the strong learning ability of JIT defect prediction models based on pre-trained models, owing to their deep and complex network structures. Consequently, during the training process, when faced with an imbalanced dataset, the models' parameters tend to favor the class with more data. This provides an explanation for why these large pre-trained models exhibit lower recall rates in the JIT defect prediction task.
\begin{itemize}
    \item Results 2: Complicated Model is sensitive to the balance of the dataset.  
\end{itemize}

\begin{table}[h]
\renewcommand{\arraystretch}{1.5} 
\tabcolsep=0.06cm
\caption {Experimental results of the different models on 
Openstack project}
\label{tab:full-finetuneres-op}
\scalebox{.6}
\par\centerline{
\small
\begin{tabular}{|c|c|c|c|c|c|}
\hline
{\bf{Model}}&  auc score &  accuracy &  precision &  recall &   F1 score  \\ \hline
DeepJIT &       0.74 &       0.61 &       0.17 &    0.76 & 0.28 \\ \hline
CC2Vec &       0.76 &      0.27 &       0.11 &    0.98 & 0.19 \\ \hline
  RoBERTaJIT &       0.79 &      0.80 &       0.32 &    0.50 & 0.39 \\ \hline
CodeBERTJIT &       0.80 &      0.80 &       0.33 &    0.56 & 0.42 \\ \hline
    GPT2JIT &       0.79 &      0.71 &       0.26 &    0.76 & 0.39 \\ \hline
 CodeGPTJIT &       0.81 &      0.76 &       0.30 &    0.70 & 0.42 \\ \hline
    BARTJIT &       0.81 &      0.71 &       0.27 &    0.77 & 0.40 \\ \hline
  PLBARTJIT &       0.81 &      0.79 &       0.32 &    0.62 & 0.42 \\ \hline
\end{tabular}}
\end{table}

\begin{figure}[h]
    \centering
    \includegraphics[width=0.44\textwidth ]{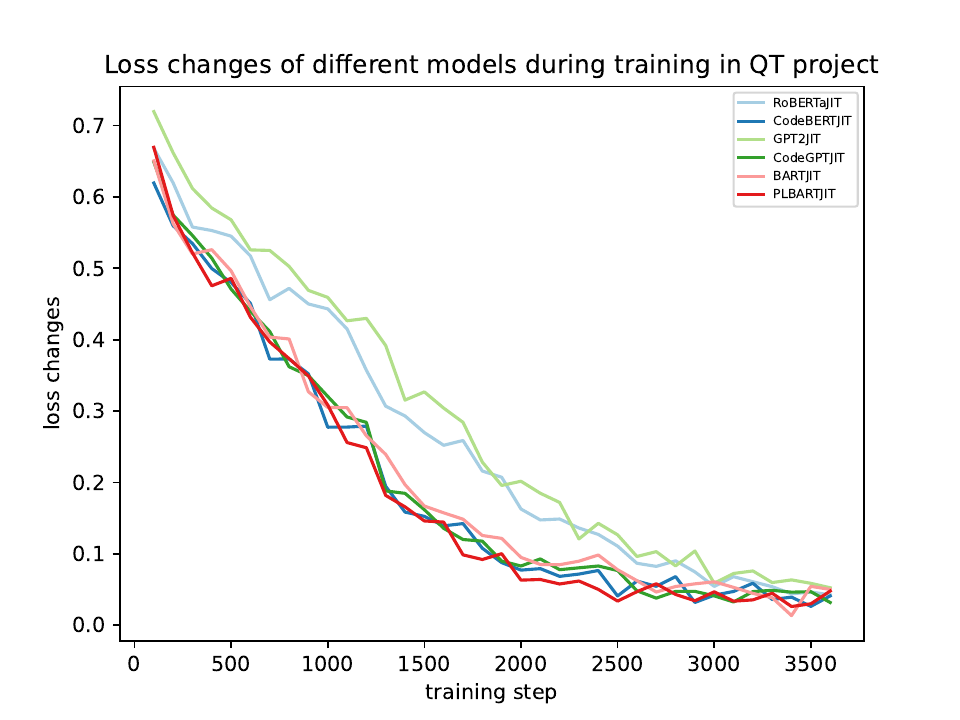}
    \caption{Loss changes during training process on QT project}
    \label{fig:QT_loss_model}
\end{figure}

\begin{figure}[h]
    \centering
    \includegraphics[width=0.44\textwidth ]{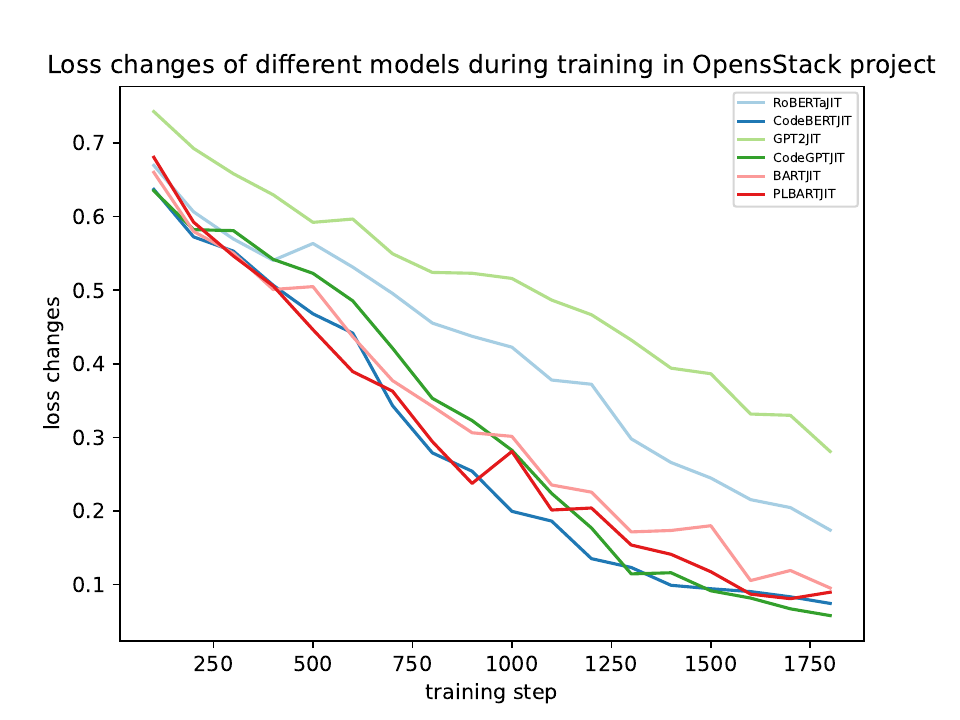}
    \caption{Loss changes during training process on Openstack project}
    \label{fig:op_loss_model}
\end{figure}

{\bfseries Results and analysis for RQ2.} We investigate the overall training efficiency of different JIT defect prediction models with pre-trained models as backbones, as well as whether there is any relationship in the distribution of prediction results among the models. We first compare the training time, parameter size, and training loss changes. Table \ref{tab:trainingtime_size} shows training time and model size respectively in RoBERTaJIT, CodeBERTJIT, GPT2JIT, CodeGPTJIT, BARTJIT, and PLBARTJIT models. Figure \ref{fig:QT_loss_model} and Figure \ref{fig:op_loss_model} respectively show training loss change of these six models in QT and Openstack projects. We also conduct T-test experiment to analyze the distribution similarity between 6 models, and Table \ref{tab:t-test} shows the T-test results between the six models on both two datasets.

Table \ref{tab:trainingtime_size} respectively shows the training time and model size of 6 models. We can find that the training time and model size of different models are not very different in general. BARTJIT and PLBART are a little faster than other models in the same training setting on both datasets. Additionally, it is evident that the training time and model size of JIT defect prediction models with similar backbone models are more comparable. For example, RoBERTaJIT is close to CodeBERTJIT, GPT2JIT is close to CodeGPTJIT, and BARTJIT is close to PLBARTJIT. Moreover, Figure \ref{fig:QT_loss_model} and Figure \ref{fig:op_loss_model} show the loss changes during the training process of the six models. It is observed that RoBERTaJIT, GPT2JIT, and BARTJIT perform slightly worse than CodeBERTJIT, CodeGPTJIT, and PLBARTJIT in terms of coverage performance. This indicates that code-related backbones provide better initial representations in the JIT defect prediction task, even though both Commit code and Commit message are treated as token types.
\begin{itemize}
    \item Results 1: The training time and model size of different JIT defect prediction models with similar backbone models tend to be closer, backbone model which is code-related converges better than their corresponding non-code-related model.
\end{itemize}

\begin{table}[h]
\renewcommand{\arraystretch}{1.5} 
\tabcolsep=0.12cm
\caption {Training time and model size of different JIT defect prediction models}
\label{tab:trainingtime_size}
\par\centerline{
\small
\begin{tabular}{|c|c|c|c|c|}
\hline
{\bf{Model}}& QT & Openstack & QT & Openstack    \\ \hline
RoBERTaJIT    &1938s    & 1005s   & 2.56 GB & 804 MB   \\ \hline
CodeBERTJIT   &1942s & 1007s   &2.57 GB& 821 MB \ \\ \hline
GPT2JIT    &2184s    & 1132s   &2.57 GB    & 815 MB   \\ \hline
CodeGPTJIT   &2150s & 1114s    &2.57 GB & 821 MB  \\ \hline
BARTJIT     &1460s    & 757s    &2.61 GB    & 863 MB     \\ \hline
PLBARTJIT   &1489s & 808s  &2.61 GB & 860 MB  \\ \hline
\end{tabular}}
\end{table}


In RQ2, we explored the relationship between different models in terms of training efficiency, but it remained unclear whether these models exhibited the same distribution in their prediction results. To investigate this further, we conducted T-test experiments between the prediction results of the six models. Table \ref{tab:t-test} presents the T-value and P-value for these models, The number on the left represents the T-value between the two models, and the number on the right represents the corresponding p-value, we keep the result to two decimal places. Symbol $\oslash$ means duplicate results that is omitted from the calculation of T-value and P-value. This suggests that despite sharing similar backbone models, the JIT defect prediction models still yield different prediction result distributions in a statistical sense.
\begin{itemize}
\item Results 2: Despite the similarity of some backbone models, the distribution of results in JIT defect prediction tasks is significantly different.
\end{itemize}

\begin{table*}
\renewcommand{\arraystretch}{1.5} 
\tabcolsep=0.18cm
\caption {T-test results for the distribution of prediction results of different models}
\label{tab:t-test}
\centerline{
\small
\begin{tabular}{|c|c|c|c|c|c|c|}
\hline
{\bf{Model}}&  RoBERTaJIT &  CodeBERTJIT &  GPT2JIT &  CodeGPTJIT &  BARTJIT&PLBARTJIT  \\ \hline
 RoBERTaJIT &    $\oslash$ &  18.41/ 0.0 &  -8.73/ 0.0 &   6.48/ 0.0 & -11.95/ 0.0 & 14.88/ 0.0\\ \hline
CodeBERTJIT &   $\oslash$ &    $\oslash$ & -29.21/ 0.0 & -13.22/ 0.0 & -33.01/ 0.0 & -3.21/ 0.001  \\ \hline
    GPT2JIT & $\oslash$ & $\oslash$ &   $\oslash$&    16.38/ 0.0 &   -3.3/ 0.001 & 25.01/ 0.0\\ \hline
 CodeGPTJIT &    $\oslash$ &  $\oslash$ &  $\oslash$ &   $\oslash$& -19.98/ 0.0 &  9.45/ 0.0\\ \hline
    BARTJIT & $\oslash$ & $\oslash$ &  $\oslash$ & $\oslash$ &    $\oslash$&  27.6/ 0.0 \\ \hline
  PLBARTJIT &    $\oslash$ &  $\oslash$ &  $\oslash$ &   $\oslash$ &   $\oslash$ &  $\oslash$\\ \hline

\end{tabular}}
\end{table*}
{\bfseries Results and analysis on RQ3.} In this research question, we aim to explore how the inclusion of Commit code and Commit message as model input affects the performance of defect detection in JIT defect prediction models. To do so, we conduct an ablation experiment where we remove either the Commit code or Commit message from the input data and analyze the impact on recall and precision, which are the most intuitive representations of defect detection and false positives.

\begin{figure*}[h]
    \centering
    \includegraphics[width=0.8\textwidth ]{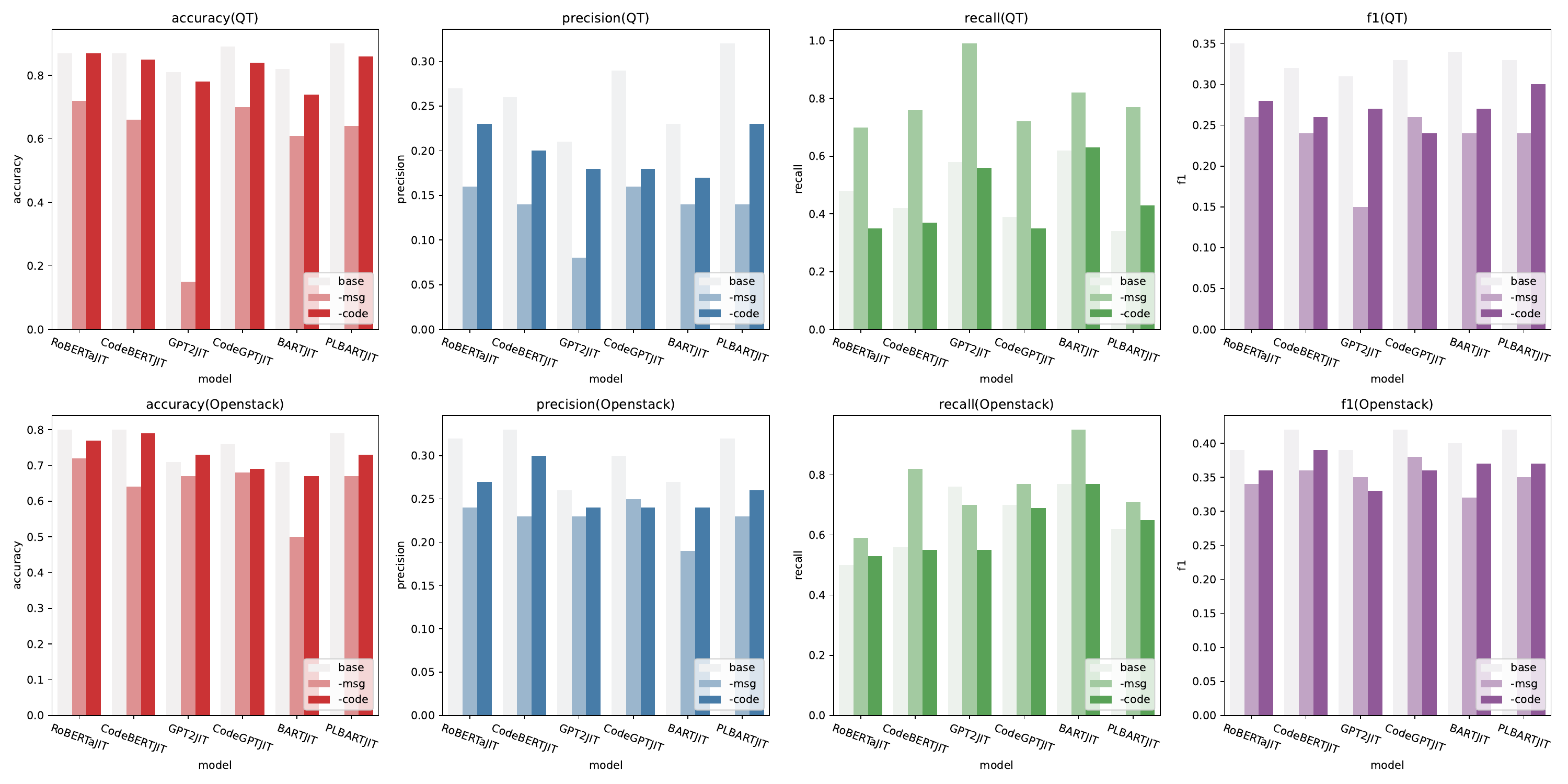}
    \caption{The results of ablation experiments of six models on two QT and Openstack datasets}
    \label{fig:ablation}
\end{figure*}

Figure \ref{fig:ablation} displays the experimental results of the six JIT defect prediction models under the ablation settings. Specifically, the ``-msg'' operation refers to using only Commit code in the model input, while the ``-code'' operation refers to using only Commit message in the model input. The ablation experiment reveals that Commit code plays a crucial role in defect detection. 

Recalls mean we correctly predict a defective submission as a defect. When we conduct the ``-msg'' operation on all six JIT defect prediction models for both QT and Openstack datasets, most of the results show a direct improvement in recall. This indicates that removing the Commit message from the input data helps the models identify more defects correctly. It is worth noting that Commit message, being provided entirely by the developer and not strictly regulated, may introduce information that does not exactly match the code description for a given commit. On the other hand, when we conduct the ``-code'' operation on the six models, more than half of the results show a decrease in recall value, while other results vary only slightly from the base model results. This further supports the finding that Commit code plays a significant role in defect detection. In addition to the main results, we made an interesting observation during the ablation experiment. Regardless of whether we removed Commit code or Commit message from the model input, the other three evaluation indicators (accuracy, precision, and F1 score) show a decline. This suggests that both parts of the data, Commit code, and Commit message, together play a crucial role in determining the overall performance of these three evaluation indicators. These findings provide valuable insights into the importance of each input data component and can guide researchers in optimizing their models for JIT defect prediction.
\begin{itemize}
\item Results: Commit code plays the main role in defect detection, both Commit message and Commit code are responsible for prediction precision, accuracy, and F1 score.
\end{itemize}

\begin{figure*}[h]
    \centering
    \includegraphics[width=0.8\textwidth ]{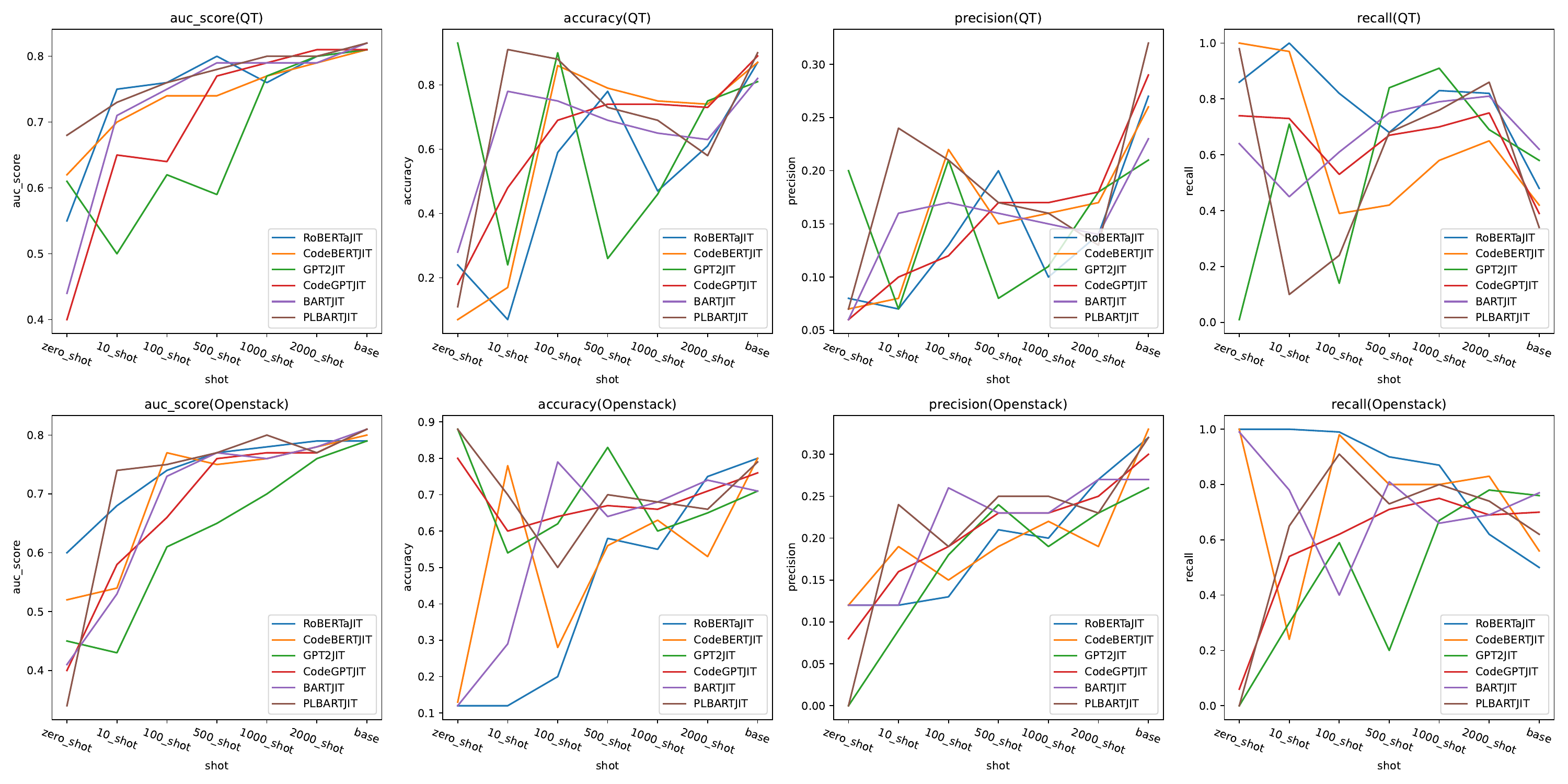}
    \caption{Model performance of different models in few-shot scenario}
    \label{fig:few-shot}
\end{figure*}

{\bfseries Results and analysis on RQ4.} In this research question, we aim to explore the performance of JIT defect prediction models with different backbone models in a few-shot scenario with a balanced dataset. To simulate real software engineering projects where defective submissions are less common, we downsampled the two datasets and set different scales (0, 10, 100, 500, 1000, and 2000) to explore the performance of the models. We used the same training setting as in RQ1 and keep the same test datasets. Figure \ref{fig:few-shot}, Table \ref{tab:2000QT}, and Table \ref{tab:2000op} present the experimental results of these 6 models in the few-shot scenarios. In the zero-shot scenario, the F1 score could not be calculated due to the problem of zero division, so we focused on comparing the model performance under the four evaluation metrics: auc score, accuracy, precision, and recall.

As shown in Figure \ref{fig:few-shot}, we observed that in the zero-shot scenario, the performance of the six models is almost abnormal, with recall and accuracy values taking on extreme values of 0 and 1. This indicates the unavailability of the models in the zero-shot scenario. However, as the dataset size gradually increases, the performance of the six models becomes more consistent and improves significantly. 
\begin{itemize}
\item Results 1: With the gradual increase in dataset size, the performance of the six models will tend to be consistent and normal. 
\end{itemize}
\begin{table}
\renewcommand{\arraystretch}{1.5} 
\tabcolsep=0.18cm
\caption {Results of QT project in 2000-shot scenario}
\label{tab:2000QT}
\centerline{
\small
\begin{tabular}{|c|c|c|c|c|}
\hline
{\bf{Model}}&  auc score &  accuracy &  precision &  recall  \\ \hline
 RoBERTaJIT &     0.81 &      0.87 &       0.27 &    0.48 \\
 2000-shot& 0.80 &      0.61 &       0.14 &    0.82\\ \hline
CodeBERTJIT &0.81 &      0.87 &       0.26 &    0.42  \\
 2000-shot&  0.79 &      0.74 &       0.17 &    0.65\\ \hline
    GPT2JIT &  0.81 &      0.81 &       0.21 &    0.58 \\ 
     2000-shot& 0.80 &      0.75 &       0.18 &    0.69\\ \hline
 CodeGPTJIT &    0.81 &      0.89 &       0.29 &    0.39 \\ 
  2000-shot&   0.81 &      0.73 &       0.18 &    0.75\\ \hline
    BARTJIT & 0.82 &      0.82 &       0.23 &    0.62\\ 
     2000-shot&  0.79 &      0.63 &       0.14 &    0.81 \\ \hline
  PLBARTJIT &    0.82 &      0.90 &       0.32 &    0.34 \\ 
   2000-shot&  0.80 &      0.58 &       0.13 &    0.86 \\ \hline
\end{tabular}}
\end{table}

Table \ref{tab:2000QT} and Table \ref{tab:2000op} show the results with 2000 training samples. From the results, we can observe interesting patterns when the dataset size shrinks to a fraction of the original datasets (0.07 times for QT and 0.15 times for Openstack). While the difference in AUC scores between the models becomes less obvious, ranging from 0 to 0.04, there is a significant difference in accuracy and recall. The significant decrease in accuracy when training the models on a much smaller dataset can be attributed to the convergence of the model. Training on a smaller dataset with the same epoch leads to a less stable convergence compared to training on larger datasets. However, despite the challenges posed by the smaller dataset size, we find that training the models on such limited data can still improve the recall rate significantly. On average, the recall rate improves by 36\% in the QT dataset and 14\% in the Openstack dataset. This finding suggests that when using a balanced dataset, JIT defect prediction models based on pre-trained models can effectively detect defects even in few-shot scenarios.
\begin{itemize}
\item Results 2: Few shot scenarios with balanced data at the right size can bring the capability of defect detection increased. 
\end{itemize}
\begin{table}
\renewcommand{\arraystretch}{1.5} 
\tabcolsep=0.18cm
\caption {Results of Openstack project in 2000-shot scenario}
\label{tab:2000op}
\centerline{
\small
\begin{tabular}{|c|c|c|c|c|}
\hline
{\bf{Model}}&  auc score &  accuracy &  precision &  recall  \\ \hline
 RoBERTaJIT &    0.79 &      0.80 &       0.32 &    0.50 \\
 2000-shot&  0.79 &      0.75 &       0.27 &    0.62\\ \hline
CodeBERTJIT &0.80 &      0.80 &       0.33 &    0.56  \\
 2000-shot&  0.78 &      0.53 &       0.19 &    0.83\\ \hline
    GPT2JIT &   0.79 &      0.71 &       0.26 &    0.76 \\ 
     2000-shot& 0.76 &      0.65 &       0.23 &    0.78\\ \hline
 CodeGPTJIT &   0.81 &      0.76 &       0.30 &    0.70  \\ 
  2000-shot&  0.77 &      0.71 &       0.25 &    0.69\\ \hline
    BARTJIT &0.81 &      0.71 &       0.27 &    0.77 \\ 
     2000-shot&  0.78 &      0.74 &       0.27 &    0.69 \\ \hline
  PLBARTJIT &    0.81 &      0.79 &       0.32 &    0.62 \\ 
   2000-shot&   0.77 &      0.66 &       0.23 &    0.74 \\ \hline
\end{tabular}}
\end{table}
We observed that some models achieved lower evaluation scores compared to the base model in this question. However, it is interesting to note that despite lower scores, CodeGPTJIT and GPT2JIT still outperformed the two state-of-the-art models, DeepJIT and CC2Vec, under the few-shot scenario. This is evident from the results presented in Table \ref{tab:2000codegpt} and Table \ref{tab:2000gpt} for CodeGPTJIT and GPT2JIT on QT and Openstack datasets, respectively. CodeGPTJIT and GPT2JIT demonstrate optimal performance on QT and Openstack datasets, respectively, indicating their effectiveness in handling few-shot scenarios. These results highlight the potential of fine-tuning based on pre-trained models in situations where training data is limited.
\begin{table}\vspace{-1.0em}
\renewcommand{\arraystretch}{1.5} 
\tabcolsep=0.08cm
\caption {Experimental result of CodeGPTJIT in 2000-shot scenario\\ on QT project}
\label{tab:2000codegpt}
\par\centerline{
\small
\begin{tabular}{|c|c|c|c|c|c|}
\hline
{\bf{Model}}&  auc score &  accuracy &  precision &  recall  & F1 score\\ \hline
DeepJIT &       0.79 &      0.73 &       0.17 &    0.70 & 0.27 \\ \hline
CC2Vec &       0.81 &      0.56 &       0.13 &   0.87 & 0.23 \\ \hline
CodeGPT & \multirow{2}*{0.81} & \multirow{2}*{0.73} & \multirow{2}*{0.18}&\multirow{2}*{0.75}&\multirow{2}*{0.29} \\ 
(2000-shot) &  &      &       &    &\\ \hline
\end{tabular}}
\end{table}\vspace{-1.5em}
\begin{table}\vspace{-1.5em}
\renewcommand{\arraystretch}{1.5} 
\tabcolsep=0.08cm
\caption {Experimental result of GPT2JIT in 2000-shot scenario \\on Openstack project}
\label{tab:2000gpt}
\par\centerline{
\small
\begin{tabular}{|c|c|c|c|c|c|}
\hline
{\bf{Model}}&  auc score &  accuracy &  precision &  recall  & F1 score\\ \hline
DeepJIT &       0.74 &       0.61 &       0.17 &    0.76 & 0.28 \\ \hline
CC2Vec &       0.76 &      0.27 &       0.11 &  0.98 & 0.19 \\ \hline
GPT2JIT & \multirow{2}*{0.76} & \multirow{2}*{0.65} & \multirow{2}*{0.23}&\multirow{2}*{0.78}&\multirow{2}*{{0.36}} \\ 
(2000-shot) &  &      &       &    &\\ \hline
\end{tabular}}
\end{table}
\section{Related Work}
Finding defects in time can reduce the cost of repairing defects and improve the quality of software. Researchers have proposed a large number of defect prediction techniques, which mainly include module-level, file-level, and change-level defect prediction based on different prediction granularity. In 2008, Kim \cite{2008Classifying} proposed a code classification method using machine learning, i.e., codes are classified into bug and bug-free categories, and the recently submitted codes are classified with or without defects. In 2013, Kamei officially named the technique predicting defects based on software change characteristics as JIT defect prediction.

JIT defect prediction can remind software developers to review their submitted code for defects as early as possible, which significantly reduces the cost of software maintenance. Most of the traditional JIT defect prediction techniques rely on manually extracted features and use different machine learning methods such as logistic regression, SVM, etc., to train a suitable classifier. Mockus et al.\cite{2000Predicting} were the first to propose the selection of manually extracted features, which they divided into different categories, such as changes in code lines and developer expertise. These features have been extensively studied in previous works\cite{Mockus2000Identifying,1979An}. Some other studies have extended the traditional features by adding characteristics like code complexity, number of changed files, and word frequency of code modules to improve code change characterization\cite{2008Classifying,7202954,2016Studying,8453123,6772130,2003Ordering,1542070}.

The general process of JIT defect prediction technology usually includes data annotation, feature extraction, model construction, and experimental evaluation. Data annotation is used to determine whether code changes introduce defects. Sliwerski et al.\cite{2005When} first proposed the algorithm to identify the changed code introducing defects in 2005, namely the SZZ algorithm. Based on the original SZZ algorithm, the researchers continue to improve to obtain more accurate code annotation\cite{2006Automatic,2017A}.

Before 2015, the research on software defects was coarse-grained. With Hinton's research in the field of deep learning\cite{Hinton1992How,8187120,6796673}, the trend of cross-fusion among various fields became more obvious. Depending on the development of deep learning technology in recent years\cite{2012ImageNet,2015Batch,2015Deep,7780459}, the proposal of various network structures provides new ideas for feature extraction. In 2015, Yang et al.\cite{7272910} proposed a method that uses deep learning technology to improve the performance of JIT defect prediction. In this study, Deep Belief Network(DBN) was used to process and map the dimensions of manually extracted features. It was the first case using deep learning in JIT defect prediction technology. In traditional JIT defect prediction technology, researchers often connect the extracted feature with a classifier to construct the model. The classifier is usually a common algorithm in machine learning, such as random forest, support vector machine(SVM), logistic regression(LR), etc.\cite{9463103,9026802}. However, all the studies mentioned above are based on traditional manual features. In recent years, with the development of natural language processing, researchers realize that automatic feature extraction can be carried out from the code and submitted information itself\cite{8816772}\cite{9284081}.

Many researchers have found that in a submission, the commit information provided by the developer is natural language, while the code is a semi-structured language and also has naturalness. These features naturally provide the relevance of JIT defect prediction techniques to natural language processing. Due to the proposed network structures such as convolutional neural network (CNN)\cite{2012ImageNet} and recurrent neural network (RNN)\cite{2014Recurrent}, feature extraction in JIT defect prediction technology is not limited to the traditional features extracted. Hoang proposed an end-to-end JIT defect prediction framework, DeepJIT. This framework uses CNN to extract features from Commit message and Commit code in a submission, and then inputs the fused features into a layer full connection for classification, so as to achieve JIT defect prediction without manually extracting features\cite{8816772}.

In 2020, Hoang\cite{2017Multilingual} proposed a distributed vector representation of code change called CC2Vec based on GRU and hierarchical attention Network. CC2Vec vector was obtained by using information of submitted logs as training labels\cite{9284081}. Both of these JIT defect prediction models using deep learning methods performed well. In view of the excellent performance of CC2Vec, Zeng et al.\cite{2021Deep} made a more detailed analysis and comparison between DeepJIT and CC2Vec, and expanded the dataset on the original basis. While confirming the performance of CC2Vec, they also analyzed its defects, and proposed a logistic regression classifier called LApredict. It is confirmed that deep learning methods still have a lot improvement to make.

At present, the research of deep JIT defect prediction is still active. The research of embedding model and pre-trained model in natural language field makes it possible to apply many new network structures and models to JIT defect prediction. In 2017, Google proposed Transformer, a neural network structure, which provides a full attention mechanism without using CNN and RNN\cite{2017Attention}. Devlin et al. proposed BERT pre-trained model using Transformer structure\cite{2018BERT}. Liu et al.\cite{2019RoBERTa} improved the BERT model, optimized its parameter selection, expanded the BERT model corpus, and proposed the RoBERTa model. CodeBERT, proposed by Feng et al., provided a bi-modal pre-trained model for both natural and programming languages, which can be used for downstream tasks in software engineering and natural language\cite{2020CodeBERT}. For better explore the ability of CodeBERT model applied in code-related task, Zhou\cite{9609166} conduces two experiments: code search and JIT defect prediction task. Some other researchers have also tried pre-trained model applied in vulnerability detection task, they rebuild vocabulary table of CodeBERT make it more suitable for obtaining tokens in the a task specific scene\cite{msr22}.

\section{Conclusions and future work}
Our work mainly focuses on the task of JIT defect prediction based on pre-trained models with different architecture. The findings in the experimental analysis showed that each model based on different backbones exhibited improvements. Models with similar backbone pre-training showed predictions with closer distributions. Commit code played a significant role in defect detection, and balanced datasets improved defect prediction ability under few-shot scenarios. These results provide new insights for optimizing JIT defect prediction tasks using pre-trained models and highlight the factors that require more attention when constructing such models.

We have also identified some future works. Based on the results of the experiments, we can find that the Commit code largely determines the defect prediction ability of all models with different pre-trained models as the backbone. This provides an idea for our future work, that is, different treatments of Commit code can be adopted in model optimization, which may bring higher benefits. And another important finding is that a balanced dataset can bring improvement in defect prediction capacity under a few-shot scenario. This means that we can further optimize our model in a few-shot scenarios.
\section*{Acknowledgment}
The work is partially supported by CityU MF EXT (project no. 9678180).

\balance

\end{document}